\documentclass{article}
\usepackage{spconf,amsmath,graphicx, hyperref,xcolor,acronym, comment}
\usepackage{colortbl}
\usepackage{hyperref}
\hypersetup{
    colorlinks=true,
    citecolor=blue,
    linkcolor=blue,
    filecolor=magenta,      
    urlcolor=blue,
}
\usepackage{spconf,amsmath}
\usepackage{multirow}
\usepackage{multicol}
\usepackage{numprint}
 
\title{ICASSP 2022 Deep Noise Suppression Challenge}
\name{Harishchandra Dubey, Vishak Gopal, Ross Cutler, Ashkan Aazami, Sergiy Matusevych, Sebastian Braun,}
\secondlinename{Sefik Emre Eskimez, Manthan Thakker, Takuya Yoshioka, Hannes Gamper, Robert Aichner}
\address{Microsoft Corporation, Redmond, USA \\
    firstname.lastname@microsoft.com}
\begin{document}
\ninept
\maketitle
\begin{abstract}
The Deep Noise Suppression (DNS) challenge is designed to foster innovation in the area of noise suppression to achieve superior perceptual speech quality. This is the 4th DNS challenge, with the previous editions held at INTERSPEECH 2020~\cite{reddy2020interspeech}, ICASSP 2021~\cite{icassp2021challenge}, and INTERSPEECH 2021~\cite{dns2021interspeech}. We open-source datasets and test sets for researchers to train their deep noise suppression models, as well as a subjective evaluation framework based on ITU-T P.835 to rate and rank-order the challenge entries. We provide access to DNSMOS P.835 and word accuracy (WAcc) APIs to challenge participants to help with iterative model improvements. In this challenge, we introduced the following changes: 
(i) Included mobile device scenarios in the blind test set; (ii) Included a personalized noise suppression track with baseline; (iii) Added WAcc as an objective metric; (iv) Included DNSMOS P.835; (v) Made the training datasets and test sets fullband (48 kHz). We use an average of WAcc and subjective scores P.835 SIG, BAK, and OVRL to get the final score for ranking the DNS models. We believe that as a research community, we still have a long way to go in achieving excellent speech quality in challenging noisy real-world scenarios. 
\end{abstract}
\begin{keywords}
Deep Noise Suppression, P.835, Perceptual Speech Quality, Personalized Noise Suppression, Speech Enhancement
\end{keywords}
\vspace{-2mm}
\section{Introduction}
\vspace{-2mm}
In recent times, hybrid work has become the ``new normal" as the number of people working remotely has increased significantly due to the COVID-19 endemic. Audio calls in the presence of background noises such as a dog barking, a baby crying, kitchen noises, neighboring talkers, in-car noises, etc., get significantly degraded in terms of quality/intelligibility of the perceived speech. This leads to increased fatigue in audio meetings. 
Deep learning-based noise suppression (DNS) has shown promising results with superior speech quality~\cite{choi2020phase, koyama2020exploring,icassp2021challenge} which is significantly better than classical approaches~\cite{malah}. Previous DNS challenges accelerated DNS research by providing a massive training dataset, real test sets, training data synthesizer, and subjective evaluation frameworks based on ITU-T P.808~\cite{naderi2020open}, and P.835~\cite{naderi2021}. Many recent papers have leveraged the DNS challenge datasets for developing DNS models~\cite{reddy2020interspeech,icassp2021challenge,dns2021interspeech}. 

The ICASSP 2022 DNS challenge focuses on personalized and non-personalized DNS for fullband audio. Personalized denoising suppresses neighboring talkers in addition to background noises. We provide fullband datasets for training personalized and non-personalized DNS models. We collected real-world test sets, developed a framework for P.835 subjective evaluation, created APIs for DNSMOS P.835~\cite{reddy2021dnsmos} and WAcc.
 We provided the development test set, DNSMOS P.835 API, and WAcc API at the beginning of the challenge, which helps participants to optimize their models. The blind test set is released 5 days before the challenge deadline, and this is used for ranking the models for DNS performance. We evaluated the submitted models based on ITU-T P.835 subjective evaluation scores, namely speech quality (SIG), background noise quality (BAK), and overall audio quality (OVRL), as well as WAcc from a state-of-the-art speech recognition system. In addition, we make DNSMOS P.835~\cite{reddy2021dnsmos} freely available to researchers. DNSMOS P.835 is a deep learning model that predicts SIG, BAK, OVRL scores for a noisy test clip. 
\vspace{-4mm}
\section{Challenge Tracks}
\label{challenge_tracks}
\vspace{-2mm}
This challenge has two tracks, namely (1) non-personalized DNS and (2) personalized DNS (PDNS) for fullband audio. Unlike previous challenges, this time, we did not have wideband (16 kHz) data in our training data and testset. Similar to previous DNS Challenges, we provide a training data synthesizer that could also be used with other datasets if participants choose to do so. The data synthesizer, configuration, scripts to access the Azure APIs, and data download scripts are provided in the challenge Github repository\footnote{\url{https://github.com/microsoft/DNS-Challenge}}.

We provided a baseline model for Track 1\footnote{\url{https://github.com/microsoft/DNS-Challenge/blob/master/NSNet2-baseline/nsnet2-20ms-48k-baseline.onnx}} and baseline enhanced test clips for Track 2. In this paper, we briefly describe the baseline models for both tracks. We adopted WAcc as an objective metric for measuring the impact of DNS on speech recognition systems. Participants submitted enhanced clips for one or both tracks. We conducted ITU-T P.835 and WAcc computation on submitted enhanced clips. The motivation to add WAcc as an evaluation metric stems from the fact that several models from past challenges had noticeable WAcc degradation resulting from over-suppression of noise and/or speech distortions. We provided the participants with an Azure API for estimating WAcc on the development set. We computed DNSMOS P.835~\cite{reddy2021dnsmos} for each audio clip in the training set and provided this to participants. DNSMOS P.835 scores can be used to segment the training dataset for conducting the data ablation studies. Participants can do experiments with different portions of the training dataset based on a chosen threshold for SIG, BAK, and OVRL. The computational requirements for the challenge tracks are described in \url{https://aka.ms/dns-challenge} 
\vspace{-4mm}
\section{Datasets}
\vspace{-2mm}
\subsection{Training Datasets}
We provide clean speech, noise, impulse responses, and a training data synthesizer for both tracks. The same noise and impulse responses are provided for both tracks. Each track has its training data synthesizer. Our training data consists of English read speech, English singing, French, German, Italian, Russian, and Spanish languages. The PDNS track has clean speech, where each audio clip consists of a concatenation of all audio clips belonging to a talker. We also provide a baseline speaker embedding for each talker in the PDNS training set. We choose clean speech with DNSMOS P.835 SIG, BAK, and OVRL to have a score of more than 4.25. The PDNS track leverages clean speech with a DNSMOS P.835 OVRL $\geq$ 4.25. Next, the audio files for each talker are combined into a single clip. We randomly sample 2.5 minutes of clean speech for each speaker and provide it as enrollment speech. We extract speaker embeddings for each enrollment audio using a baseline RawNet2 speaker model~\cite{jung2019rawnet} which is adopted as the baseline speaker embedding extractor for this challenge. The next sections describe the clean speech and noise data.
\vspace{-2mm}
\subsubsection{Clean Speech}
\label{ssec:cleanspeech}
\vspace{-2mm}
The clean speech consists of six languages, namely English, French, German, Italian, Russian, and Spanish. English clean speech consists of read speech and singing, while the rest of the languages only have read speech. We provide DNSMOS P.835 scores to help participants filter the data based on DNSMOS P.835 scores if they want to do data ablation studies. The personalized track consists of clips with DNSMOS P.835 OVRL $\geq$ 4.25. We combine all audio clips from each unique talker into a single file. The PDNS training data has a total of 3230 talkers, out of which 60\% of the talkers are randomly chosen to be primary talkers while the rest are neighboring talkers. We provide the file list for PDNS clean speech with `primary' and `secondary' tags. Challenge participants can use the provided primary/secondary tags or generate their own. We sampled 60\% of the speakers as primary talkers, ensuring a uniform distribution among all languages, read speech or singing. 

English clean speech is derived from Librivox\footnote{\url{https://librivox.org/}} where we include audio clips chosen using a subjective ITU-T P.808 framework~\cite{naderi2020open}. English singing data consists of high-quality audio recordings from professional singers contained in the \textit{VocalSet} corpus~\cite{wilkins2018vocalset}. It has 10.1 hours of clean singing recorded by 20 professional singers: 9 males, and 11 females. This data was recorded on a range of vowels, a diverse set of voices on several standard and extended vocal techniques, and sung in contexts of scales, arpeggios, long tones, and excerpts. The PDNS English clean speech contains 1934 talkers from Librivox, 110 talkers from VCTK corpus, 20 talkers from Vocalset. The PDNS clean speech consists of 47 talkers for French, 874 for German, 14 for Italian, 7 for Russian, and 224 for Spanish. 
\begin{figure}[!tb]
\centering
\includegraphics[width=0.7\columnwidth]{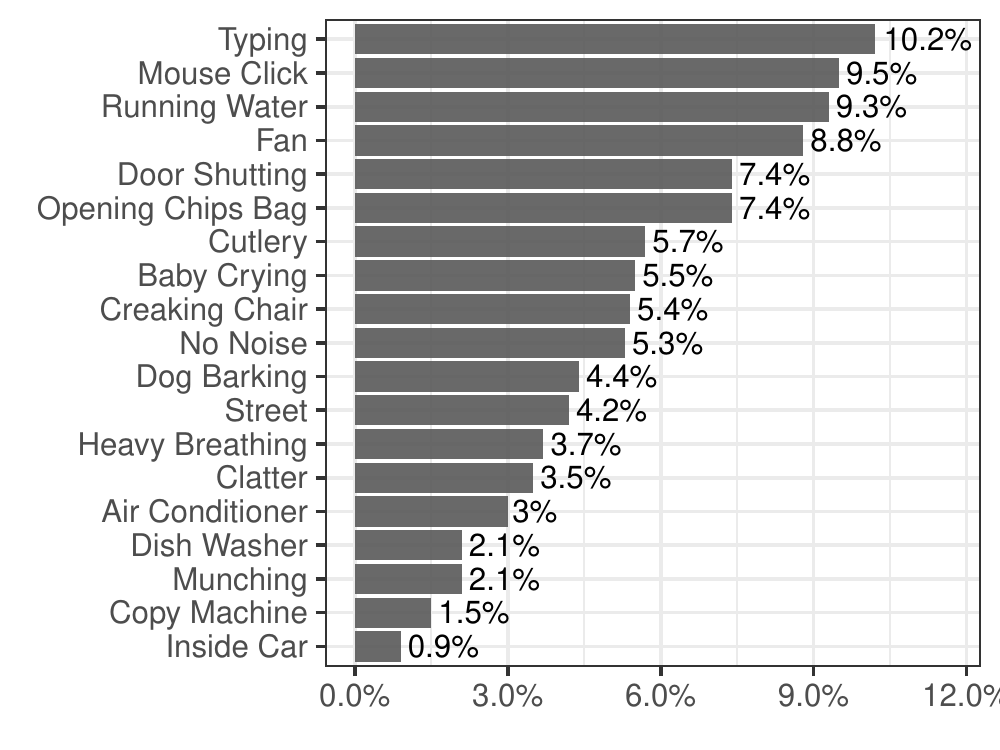}
\vspace{-4mm}
\caption{Distribution of noise types in our blind test set.}
\label{fig:noisetype}
\end{figure}
\vspace{-1mm}
\subsubsection{Noise}
\label{ssec:noise}
 We use the same noise clips for both tracks. Noise data consists of about 62,000 clips belonging to 150 noise classes. The noise clips were chosen from Audio Set\footnote{\url{https://research.google.com/Audio Set}}~\cite{7952261} and Freesound\footnote{\url{https://freesound.org/}}. Audio Set is a collection of about 2 million human-labeled 10s sound clips drawn from YouTube videos belonging to about 600 audio events. Certain audio event classes are over-represented in Audio Set. For example, there are over a million clips with audio classes music and speech and less than 200 clips for classes such as toothbrush, creaking, etc. Approximately 42\% of the clips have a single class, but the rest may have 2 to 15 labels. We developed a sampling approach to balance the noise classes in our dataset such that each class has at least 500 clips. We used a speech activity detector to remove the clips with any kind of speech activity (voice content) from our noise clips. This enabled us to get noise clips with no presence of speech. We augmented the Audio Set noise clips with 10,000 noise clips downloaded from Freesound and DEMAND databases~\cite{demand}. The total noise data constitute 181 hours of audio. Fig.~\ref{fig:noisetype} shows the histogram of noise classes included in the blind test set. These noise types were validated by a human listener after the collection of the blind set.
\vspace{-1mm}
\subsubsection{Impulse Responses}
\vspace{-1mm}
We provide 248 real and about 60,000 synthetic room impulse responses, which can be leveraged for generating reverberant noisy training data. The training data synthesizer adds noise to reverberant clean speech depending on the chosen configuration. Participants may choose to use clean speech or reverberant speech as training targets for their DNS models. We chose impulse responses from the openSLR26 and openSLR28~\cite{ko2017study} datasets.
\vspace{-2mm}
\subsection{Test set}
\vspace{-1mm}
We have two test sets: dev test set and blind test set. The dev test set is intended for model development and optimization, and is provided at the start of the challenge. The blind test set is used for ranking the challenge model in terms of evaluation metrics and is intended to be used as an unseen test set. Good performance of a model on the blind test set will show it is generalized. Both test sets consist of fullband audio clips recorded in real-world scenarios collected through crowd-sourcing where workers read provided text prompts and record their voice using desktop/laptop/mobile devices in the presence of noise and/or neighboring talkers. 
\vspace{-2mm}
\subsubsection{Non-personalized Development Test set}
\label{ssec:track1}
\vspace{-2mm}
The development test set for the non-personalized track consists of 930 real recordings. All clips contain noisy speech in the English language. Among these, 193 test clips have emotional speech in the presence of noise. There are six emotion types, namely happy, sad, angry, yelling, crying, and laughter. Crowd-sourced workers were asked to read provided text prompts and create emotional events in each test clip. The remaining clips contain the voice of a talker reading text in the presence of the following noise types: fan, air conditioner, typing, door shutting, clatter noise, car noise (i.e., standing near a car on a busy street or standing outside the car), kitchen noise (noise from kitchen utensils, dish scrubbing etc.), dish washer, running water, opening chips bags, munching or eating, creaking chair, heavy breathing, copy machine, baby crying, dog barking, inside-car noise (e.g., sitting on a passenger seat in a car which is being driven by someone else), mouse clicks, mouse scroll wheel, touch pad clicks, etc. Each test clip was recorded at 48 kHz with a duration of 10 to 20 seconds. Workers were asked to record in the near-field (close-talk) and far-field with distances of 1, 2, and 3 meters. All test clips in the non-personalized development test set were recorded using a laptop or desktop computer.
\vspace{-3mm}
\subsubsection{Personalized Development Test set}
\label{ssec:track2}
\vspace{-2mm}
Both development and blind test set have 2.5 minutes of enrollment speech for primary talkers to be used in personalized denoising. PDNS leverages speaker embedding (features) for preserving only the primary talker in a noisy environment while suppressing the neighboring talkers and noise. The development test set for the personalized track consists of 1443 real recordings. All clips contain noisy speech in the English language. Among these, 193 test clips have emotional speech in the presence of noise and are identical to the emotional test clips in the non-personalized track. There are 737 test clips where the primary talker reads the provided text in the presence of the same noise types as those in the non-personalized track. Each test clip was recorded at 48~kHz with a duration of 10--20 seconds. Workers were asked to record in the near-field (close-talk) and far-field with distances of 1, 2, and 3 meters. There are 166 test clips with the primary talker speaking in the presence of a neighboring talker and noise where both the noise and neighboring talker are simultaneously active in the primary talker's background. There are 347 test clips where the primary talker is speaking in the presence of a neighboring speaker with no background noise. Thus, we have simulated three scenarios for PDNS: (i) primary talker in the presence of noise; (ii) primary talker in the presence of neighboring talker; and (iii) primary talker in the presence of simultaneously active neighboring talker and noise. All test clips in the personalized development test set were recorded using a laptop or desktop computer.

\subsection{Blind test set}

We collected a common blind test set for both tracks, which facilitates a direct comparison of both tracks to elucidate the benefits of personalized noise suppression. Track 2 leverages 2.5 minutes of clean speech enrollment data for personalized denoising. The blind set has 2.5 minutes of enrollment speech for each test clip, which is intended for use only by personalized models. There are 859 real test clips, each with a duration of 10s, in the blind test set. We collected the blind test set on a variety of desktop and mobile platforms through crowd-sourcing, where 70\% of the test clips were collected on a smartphone. The blind test set went through several iterations of data validation based on unit tests and human listening. We did not include test clips from the same speaker if they had a similar background noise. In this way, we have a unique speaker and background noises in each test clip. We transcribed the blind test set using a third-party data annotation service. We did expert listening to correct the speech transcription for the blind test set to ensure high accuracy. Participants were provided with the blind test set 5 days before the data submission deadline.

\vspace{-1mm}
\section{Results~\& Discussions}
\vspace{-2mm}
\subsection{Baseline for Non-personalized DNS}

We trained NSNet2~\cite{dns2021interspeech,braun2020data} on a subset of the fullband non-personalized training dataset to obtain the baseline. The subset was characterized by clean speech clips having the DNSMOS P.835 scores greater than or equal to 4.2, 4.5, and 4.0 for SIG, BAK, and OVRL, respectively. We denoised the Track 1 dev test-set and blind test-set with the trained NSNet2 baseline model, to get the baseline enhanced clips. We release the ONNX\footnote{\url{https://onnx.ai}} model and inference script for NSNet2. 
\vspace{-2mm}
\subsection{Baseline for Personalized DNS}
\vspace{-1mm}
The baseline for the PDNS track consists of the baseline speaker embedding extractor and the baseline PDNS model. The baseline speaker embedding extractor is RawNet2~\cite{jung2020improved} trained on wideband audio from VoxCeleb2~\cite{chung2018voxceleb2}. Lack of fullband audio datasets with speaker IDs and thousands of speakers led us to select VoxCeleb2, which is a wideband audio dataset with ~6,000 speakers. Thus our baseline PDNS model had a wideband speaker embedding extractor. We used the baseline speaker embedding extractor on all enrollment data in the training set, and provided the participants with speaker embeddings. This lowered the barrier to entry in the PDNS track. Participants were permitted to retrain the RawNet2 model with fullband data. The challenge permitted the use of external datasets in addition to DNS challenge datasets. Participants could leverage other state-of-the-art speaker models~\cite{wan2018generalized,snyder2018x,jung2020improved} for extracting speaker embeddings~\cite{wan2018generalized,snyder2018x}.

PDNS models are trained to suppress neighboring talkers and background noises and only preserve the enhanced speech from the primary talker. To achieve this, PDNS models leverage speaker embedding features extracted from the enrollment speech along with spectral features (or raw-waveform) of the noisy input audio. We used the personalized DCCRN (pDCCRN) model described in~\cite{eskimez2021personalized} as the baseline PDNS model for this challenge. We modified pDCCRN to accept 48 kHz input and added a layer to the encoder and decoder. Since the baseline speaker embedding extractor uses wideband audio, the input audio is downsampled to 16 kHz before extracting the speaker embeddings. 

Each unique talker in the personalized training dataset, PDNS dev test set, and blind test set has 2.5 minutes of enrollment speech. PDNS models are expected to leverage talker-aware training and talker-adapted inference. There are two motivations to provide clean speech for enrollment of the primary talker: (1) speaker models are sensitive to false-alarms in speech activity detection (SAD)~\cite{hansen2015speaker}; clean speech can be used for obtaining accurate SAD labels resulting in speaker-discriminative embeddings. (2) Speaker adaptation is expected to work well using multi-conditioned data; clean speech can be used for generating reverberant and noisy multi-condition enrollment data for speaker adaptation. 
\begin{figure}[!tb]
    \centering
    \includegraphics[width=0.6\columnwidth]{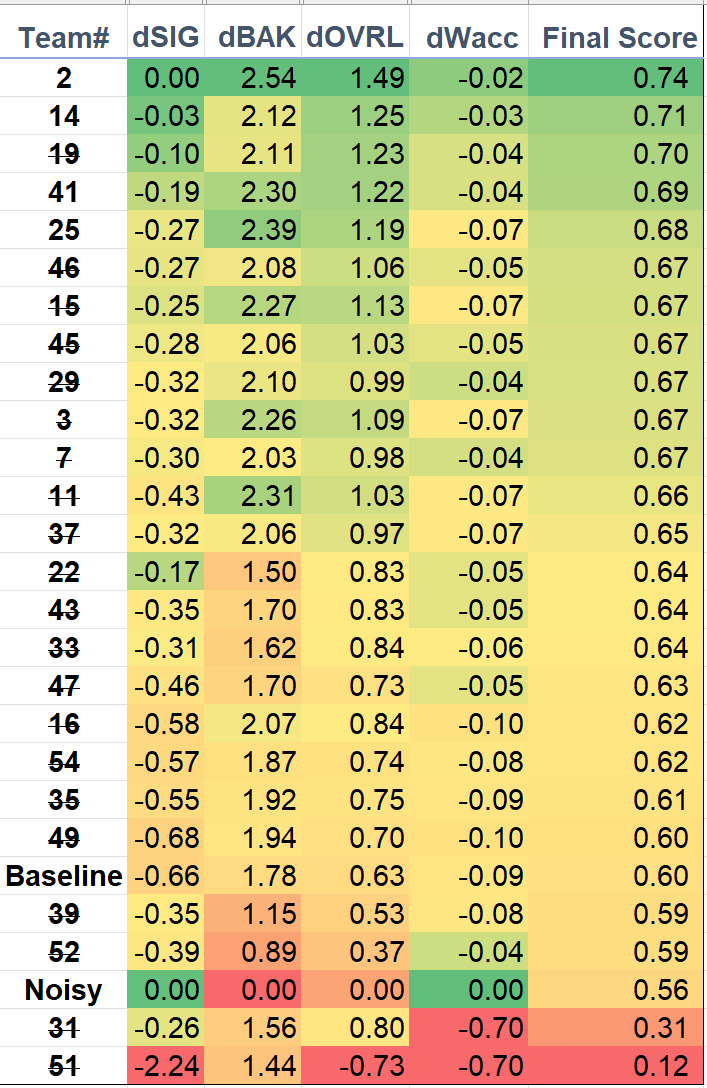}
    \caption{Results: P.835 subjective evaluation of all models from Track 1 non-personalized DNS on blind testset.}
    \label{fig:p835}
\end{figure}
\begin{figure}[!tb]
    \centering
    \includegraphics[width=0.6\columnwidth]{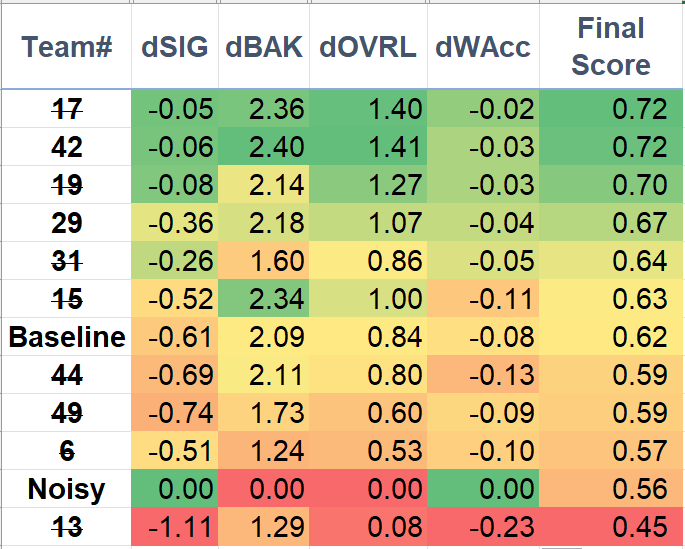}
    \caption{Results: P.835 subjective evaluation of all models from Track 2 personalized DNS on blind testset.}
    \label{fig:p835_pdns}
    
\end{figure}
\vspace{-2mm}
\subsection{Evaluation Methodology}
\vspace{-2mm}
This challenge relies on ITU-T P.835~\cite{naderi2021} subjective evaluation for evaluating the DNS models, since objective speech quality metrics, such as PESQ~\cite{p862}, SDR, and POLQA~\cite{polqa}, do not correlate well with subjective speech quality~\cite{Avila2019}. A modified version of ITU-T P.835 was used for measuring the performance of personalized models. The modified P.835 for personalized DNS provides 10s of enrollment speech of the primary speaker so that the raters can recognize the primary talker's voice while assigning subjective scores. Human raters were instructed to focus on the quality of the voice of the primary talker when more than two talkers were present in a test clip. 

Four metrics, three P.835 subjective scores (SIG, BAK, OVRL) and WAcc from a speech recognition system were used to evaluate challenge models. Scores on the blind test set were combined into a final score for ranking the models. Higher WAcc shows superior denoising performance with respect to speech recognition. WAcc is defined as $\text{WAcc} = 1 - \text{WER}$, where WER is the word error rate of speech recognition system. The final score is computed as\\ $\text{Final score}= 0.5[\text{WAcc} + 0.25(\text{OVRL} -1)]$.

%
\vspace{-2mm}
\subsection{Results}
\vspace{-2mm}
We received 24 and 10 submissions for Track 1 and Track 2, respectively. Each team submitted a processed blind test set (Sec.~3.3).

Fig.~\ref{fig:p835} and Fig.~\ref{fig:p835_pdns} show the subjective P.835 scores, WAcc and final score for challenge entries in decreasing order of performance. dSIG, dBAK, dOVRL refers to the difference in SIG, BAK, OVRL between the enhanced clip and noisy clip. Similarly, dWAcc is the difference in WAcc between the enhanced clip and noisy clips. 


For the top performing teams, we ran an ANOVA test to determine statistical significance (see \url{https://aka.ms/dns-challenge}). The 2nd, 3rd and 4th place are tied for Track 1. Likewise the 1st and 2nd place for Track 2 are tied. Teams 17, 19, and 42  did not submit a paper so were disqualified per the challenge rules.

From a breakdown of scores based on the device type (mobile/desktop) we find that the MOS scores for clips recorded on mobile devices is higher than those from desktop devices (see \url{https://aka.ms/dns-challenge}). This suggests that mobile had better acoustic devices or environments than the desktop scenarios.

 We required participants to not do any automatic gain control (AGC). Also, we did not perform any AGC on blind test clips or enhanced clips. A state-of-the-art speech recognition API from Azure Cognitive service was used for computing WAcc. The speech recognition system was trained to handle audio with a wide range of energy levels so we do not expect any degradation of WAcc due to varying energy levels in clips from the blind test set. Table \ref{tab:dnsmos_coeff} shows the Pearson correlation coefficient (PCC) and Spearman’s rank correlation coefficient (SRCC) between per-model subjective scores and corresponding DNSMOS P.835 scores~\cite{reddy2021dnsmos}. The high correlation between subjective scores and DNSMOS P.835 in both tracks shows the efficacy of DNSMOS P.835 in ranking the DNS models. It validated our approach for providing a dev test set and DNSMOS P.835 to challenge participants for model development. 
 
 Table \ref{tab:top_models} gives a high-level comparison of the top performing models. Note there is low correlation of model size or real-time factor with performance. The top performing models for Track 1 didn't use additional datasets, while the Track 2 models did. The winning team for Track 1 \cite{zhang_multi-scale_2022} also won the ICASSP 2022 AEC Challenge \cite{cutler_icassp_2022}, and demonstrates a single model can provide excellent AEC, DNS, and WAcc performance. The performance of the personalized DNS track also show excellent performance, greatly exceeding results of our first personalized DNS challenge \cite{icassp2021challenge} with the winner \cite{Team42_Meet_TEA} providing very good dOVRL with low dSIG and low dWAcc. Note, however, that no team actually improves SIG and WAcc is 2\% worse than noisy. There is still a lot of room for improvement.  
\begin{table}[t!]
    \centering
    \caption{DNSMOS PCC and SRCC}
    \label{tab:dnsmos_coeff}
    \vspace{1mm}
    \begin{tabular}{ |c||c|c|c||c|c|c| }
    \hline
    \rowcolor{lightgray} \multirow{2}{2em}{ } & \multicolumn{3}{c||}{Track 1} & \multicolumn{3}{c|}{Track 2} \\
    \cline{2-7}
        \rowcolor{lightgray} &  SIG & BAK & OVRL & SIG & BAK & OVRL \\
    \hline
        PCC & 0.93 & 0.92 & 0.94 & 0.92 & 0.96 & 0.96 \\
    \hline
        SRCC & 0.78 & 0.89 & 0.85 & 0.84 & 0.89 & 0.93 \\
    \hline
    \end{tabular}
\end{table}
\begin{table}[t!]
    \centering
    \caption{Comparison of top performing models.}
    \label{tab:top_models}
    \begin{tabular}{|c|c|r|p{3em}|p{4em}|}
    \hline
    \rowcolor{lightgray} \textbf{Track} & \textbf{Team} & \textbf{Params} & \textbf{Real-time Factor} & \textbf{Additional data-sets} \\
    \hline
        1 & 2~\cite{Team2_baidu} & 1.5M & 0.60 & N \\
        \hline
        1 & 14~\cite{Team14_Alibaba_NTU} & 10.27 M & 0.68 & N \\
        \hline
        1 & 41~\cite{Team41_Harbin} & 29.9 M & 0.45 & N \\
        \hline
        1 & 25~\cite{Team25_CMRI_BJTU} & 5.29 M & 0.65 & N \\
        \hline
        2 & 42~\cite{Team42_Meet_TEA} & 7.81 M & 0.96 & Y \\
        \hline
        2 & 29~\cite{Team29_Kuaishou} & 12.41 M & 0.19 & Y \\
        \hline
    \end{tabular}
\end{table}
\vspace{-2mm}
\section{Conclusion}
\vspace{-1mm}
We hope this challenge dataset, test set, test framework, DNSMOS P.835, and top performing papers (Table \ref{tab:top_models}) help push the field forward. The next DNS challenge will have a more diverse test set including more languages, accents, and devices from realistic noisy scenarios. We plan to have a dedicated inference engine/evaluation setup for computing the model complexity and inference time for all submitted models. We will also include a validation of the look-ahead to make sure the comparison is fair for all models. 

%
\vfill\pagebreak
\bibliographystyle{IEEEbib}
\bibliography{strings,refs,IC3-AI}
\end{document}